\documentstyle[11pt,aasms4]{article}
\def\la{\hbox{\raise.35ex\rlap{$<$}\lower.6ex\hbox{$\sim$}\ }}
\def\ga{\hbox{\raise.35ex\rlap{$>$}\lower.6ex\hbox{$\sim$}\ }}
\def\beq{\begin{equation}}
\def\eeq{\end{equation}}
\def\beqa{\begin{eqnarray}}
\def\eeqa{\end{eqnarray}}

\def\amp{A}
\def\u{{\bf u}}
\def\sub#1{_{_{#1}}}

\righthead{Umurhan, Tao \& Spiegel: Stellar Oscillons}
\lefthead{Annals of the New York Academny of Sciences}

\begin{document}

\title{Stellar Oscillons}

\author{O.M. Umurhan, \  L. Tao and E.A. Spiegel}
\affil{\it Department of Astronomy\\ Columbia University \\
New York, NY 10027}

\vskip 20pt


\begin{abstract}
We study the weakly nonlinear evolution of acoustic instability
of a plane-parallel polytrope with thermal dissipation in the form of
Newton's law of cooling.  The most unstable horizontal wavenumbers form a
band around zero and this permits the development of a nonlinear pattern
theory leading to a complex Ginzburg-Landau equation (CGLE).  Numerical
solutions for a subcritical, quintic CGLE produce vertically
oscillating, localized structures that resemble the oscillons observed in
recent experiments of vibrated granular material.
\end{abstract}


\section{Introduction}
The excitation of sound waves in the envelopes of stars has been
extensively studied for its diagnostic importance as well as for its
intrinsic physical interest.  The most familiar mechanism of stellar
acoustic instability is the Eddington valve mechanism or
kappa mechanism, so-called because it relies on the dependence of opacity,
$\kappa$, on physical conditions.  This mechanism, which is basically
thermal, resembles the phenomenon of negative differential resistivity$^1$
familiar in condensed matter physics.  However, sound waves
can become unstable even when there is no kappa mechanism operating, and
we here discuss a simple version of such instability.

The case of optically thin perturbations to a fluid layer stratified
under gravity is one where the kappa mechanism cannot operate and yet it
does show instabilities of sound waves under suitable conditions$^2$.
The space available for this work is not sufficient
for a discussion of the conditions under which such instabilities can
occur (but see Umurhan$^3$).   Moreover this paper is a contribution
to a symposium on nonlinear astrophysics, so our aim is to describe
some nonlinear aspects of these acoustic instabilities, with pauses on the
way for only a few indispensable remarks about the linear theory in
section 3, following an introduction to the basic equations in section
2.  In section 4 we outline the nonlinear procedures and conclude in
section 5.

\section{Equations and Equilibria}
We consider the dynamics of a plane-parallel fluid subject to some
form of radiative heat exchange but not to viscosity.  The equations
of motion for this system are,
\beqa
\partial\sub t \rho + \nabla\cdot \left(\rho \u\right) & = & 0
\label{continuity} \\
\rho\left(\partial\sub t + \u\cdot \nabla\right) \u
& = &  -\nabla p + \rho g \bf{{\hat z}}
\label{momentum} \\
C\sub v\rho(\partial\sub t + \u\cdot\nabla) T +
p\nabla\cdot \u & = & Q(T,\rho) \label{heat} \\
p & = & {\cal R}\rho T \label{eqnofstate}
\eeqa
where the variables $\rho$, $\u$, $T$ and $p$ are respectively
density, vector velocity, temperature and pressure and $Q(T,\rho)$
represents the thermal sources and sinks of the medium due to radiation
and possibly mechanical effects.  The vertical coordinate $z$ is measured
positively downward and $\hat{\bf z}$ is a downward-pointing unit vector.

In equilibrium ${\bf u}={\bf 0}$ and the state variables depend only on
$z$.  The governing equations are the hydrostatic equation, the equation
of state, and the thermal equilibrium condition, $Q(T_0, \rho_0)=0$, where
the subscript naught denotes equilibrium values.  Rather than go into the
details of the transfer problem in the equilibrium state, we simply
postulate that there is an equilibrium in which $T_0(z)=\beta z$,
where $\beta$ is a constant; this is in fact the state one obtains
in the diffusion limit with no heating and with a fairly general form for
the opacity$^4$.  With the hydrostatic condition and the
equation of state, we then find that $\rho_0(z) = \rho_* (z/z_*)^m$ and
$p_0(z) = p_* (z/z_*)^{m+1}$ where $\rho_*$ and $p_*$ are constants
and $m=g/({\cal R}\beta)-1$ is an atmospheric analogue of the polytropic
index.  We let $T_*=\beta z_*$, and choose $p_*=p_0(z_*)$,
$\rho_* = \rho_0(z_*)$, so we have $p_* = {\cal R}\rho_*T_*$.

We introduce natural units so that there remain only nondimensional
equations in evidence.  We let $z_*$ be the unit of length, the speed of
sound ($c_a=\sqrt{\gamma{\cal R}T_*}$) be the unit of speed, $\rho_*$
be the unit of density and so on.  The equilibrium temperature is
then $T_0=z$ and similarly for the other quantities.

We assume that the atmosphere is truncated so that the equilibrium
thermodynamic quantities are nowhere zero.   The fluid is confined
between $z=1$ below and $z=z_0>0$ above, where $(1- z_0)$ is the
nondimensional layer thickness.  We require that the vertical velocity
vanishes on $z=z_0$ and $z=1$.

\section{Linear Theory}
We now introduce small perturbations about the static basic
state and linearize the resulting equations to study the stability
of the equilibrium state.  We shall consider a very simple version
of the stability theory here since our aim is to bring out features
of the nonlinear aspects.

The most complicated physical issue is the treatment of the
transfer problem in the general case.  For weak, optically thin
perturbations to the equilibrium, we have that
\begin{equation}
Q(T, \rho) = Q(T_0, \rho_0) + Q_T(T_0, \rho_0) (T - T_0) +
Q_\rho (T_0, \rho_0) (\rho - \rho_0) + ... \label{Q}
\end{equation}
where the subscripts represent differentiation.  In Newton's law
of cooling, $Q_\rho=0$ and the higher order terms are neglected. We
adopt that form here and write $Q_T = - \rho_0 C_v q$, where $q$ is
a characteristic inverse time.

For a grey, optically thin medium, we may express $q$ in terms of the
absorption coefficient and the state variables$^5$.  If we
assume that this coefficient is proportional to a power of density times a
power of temperature, the linear theory is straightforward and acoustic
instabilities occur in several parameter regimes$^{2,6,7}$. In particular,
the case of constant $q$ is very simple and we shall adopt it here and
write for the rest of this discussion that $Q(T, \rho) = -q \rho\sub 0
C\sub v(T-T_0)$ with $q$ constant.

If we expand about the equilibrium solution and linearize, we obtain
a set of equations that is tractable both analytically and numerically.
These linear equations are separable and the general perturbation is
written in the form \begin{equation}
f(x,y,z,t)=F(z)\exp[\sigma t + i (\omega t + {\bf k}\cdot {\bf x})]
\label{separation} \end{equation}
where ${\bf k}$ is the horizontal wave vector and $\sigma$ and
$\omega$ are real. This leads to a confluent hypergeometric equation if
suitable dependent variables are introduced$^2$.

As usual, we find gravity (or convective) modes and acoustic modes, but
we focus only on the fundamental acoustic mode for the purposes of this
discussion. In Figure 1 we plot $\omega$ and $\sigma$ vs.\ $k= |{\bf k}|$
for the case of $\gamma=1.28$, $m = 1.5$, $z_0 = 0.1$ and
$q = 3$.  We see that there is instability in a band
of wavenumbers around zero.  From this band, we then construct a
nonlinear wave packet in the next section.

\section{Acoustic Pattern Equations}
The linear problem is characterized by a band of overstable
wavenumbers around $k=0$ and we have what is called a
Hopf bifurcation in nonlinear stability theory.  In this circumstance, the
generic equation governing the nonlinear spatio-temporal evolution of a
wave packet is a complex Ginzburg-Landau equation.  To derive this
equation, we use a multiple-scale analysis based on the availability of
a small parameter, here the degree of instability of the layer.
\par
The linear theory provides a condition on the parameters
$q$, $m$, $\gamma$, and $z_0$ for the onset of acoustic instability.
We may fix three of these and treat the fourth, say $\gamma$, as the
control parameter governing the degree of instability.  Thus, for fixed
$m$, $z_0$ and $q$ we find that instability begins as $\gamma$
passes below the critical value $\gamma_c$ and we examine values
\beq
\gamma = \gamma_c - \epsilon^2 \mu,   \label{gam}
\eeq
where $\mu$ is simply a fiducial quantity.
It is $\epsilon$ that measures the degree of instability and
we take it to be small here.

The band of significantly unstable wavenumbers around zero has a width
of order $\epsilon$ and so we use this parameter to rescale the horizontal
coordinate in the usual manner of nonlinear instability theory$^8$.
Similarly, we scale the time and we then
look for solutions in terms of the scaled variables.    In
particular, the deviations from equilibrium temperature, density
and vertical and horizontal velocity are of the forms
\beqa
\theta & \sim & \epsilon\left[A (\epsilon x, \epsilon y,
\epsilon^2 t) \Theta(z) e^{i
\omega t} + c.c.\right] + ...\cr
\rho & \sim & \epsilon\left[A (\epsilon x, \epsilon y,
\epsilon^2 t) P(z) e^{i \omega
t} + c.c.\right] + ...\cr
w & \sim & \epsilon\left[A (\epsilon x, \epsilon y,
\epsilon^2 t) W(z) e^{i \omega t}
+ c.c.\right] + ...\cr
u & \sim & \epsilon^2 \left[A (\epsilon x, \epsilon y,
\epsilon^2 t) U(z) e^{i \omega
t} + c.c.\right] + ..., \label{pert}
\eeqa
where $A$ is the (generally complex) envelope function that describes the
pattern of the instability.  In linear theory, $A$ is an arbitrary
constant; in nonlinear theory it is a slowly varying function that is the
focus of interest.  The functions of $z$, on substitution and
suitable asymptotic development, turn out to be the $z$-dependent parts
of the linear eigenfunctions.

The asymptotic developments based on these scalings leads to an
equation for $A$.  On general grounds, an overstability of the kind
we have here is known to lead to an equation of the form$^8$ \beq
\partial\sub t \amp =  \mu \amp  + \alpha \Delta \amp
+ {\cal F}(|A|^2) A \label{gen}
\eeq
where the Laplacian operates in the two dimensional space $(\epsilon x,
\epsilon y)$.  The linear part of this equation describes the linear
stability theory while the quantity ${\cal F}(|A|^2)$
represents the renormalization of the linear growth rate by nonlinear
effects, with ${\cal F}(0) = 0$.

Determination of ${\cal F}$ is not possible in general, so it is
represented by Taylor series and the asymptotic development allows the
computation of the coefficients in this development at each order in
$\epsilon$.  Typically, only the leading order is needed for weak
instability.  Then equation (\ref{gen}) becomes the (cubic)
complex Ginzburg-Landau equation (CGLE), which generally describes
the patterns resulting from overstable systems with a continuous
spectrum of horizontal wave numbers.  The cubic term is
able to saturate the linear growth when $Re(\beta) < 0$.  However, if the
leading nonlinearity, the cubic term, has a coefficient that does not
allow for nonlinear saturation of the instability, higher order terms
must be sought.  This may require a modification of the scaling.

For situations without a strong symmetry in the vertical, nonlinear
saturation by the cubic term often does not occur and we have what is
called subcritical bifurcation.  This resembles a phase transition of
the first kind and it is what we see at the larger values of $q$.  In
that case, the acoustic pattern equation is of the form \beq
\partial\sub t \amp =  \mu \amp  + \alpha \Delta \amp
+ \beta |\amp|^2 \amp  + \eta |A|^4 A \ .
\label{eq:CGLE}
\eeq
The complex constants $\alpha$, $\beta$
and $\eta$ are functions of the physical
parameters of our model atmosphere.  Calculation of the coefficients
requires a working out of the nonlinear perturbation theory and we
indicate some results of such work for selected values in the table below.

\bigskip
\bigskip

\begin{tabular}{|c|c|c||l|l|l|l|l|r|}  \hline\hline
 $m$ & $z_0$ & $q$ & $\gamma_c$ &
$\alpha_r$ & $\alpha_i$ & $\beta_r$ & $\beta_i$ & Comments \\ \hline\hline
$1.5$ & $0.1$ & $3.0$ & $1.38$ & $+0.0173$ & $-0.0661$ & $+1.107$ &
$+4.754$ & subcritical \\
$6.0$ & $0.1$ & $1.0$ & $1.238$ & $+0.0003$ & $-0.041$ & $-166.1$ &
$+35.94$ & supercritical \\
\hline
\end{tabular}
\bigskip
\bigskip

Both subcritical and supercritical instabilities may occur and
produce differing behavior.  For the case of one-dimensional patterns,
spatio-temporal disorder is the rule in the supercritical case$^{2,9}$.
But in the subcritical
state, stable isolated structures may be expected, as argued
by Thual and Fauve$^{10,11}$.
In the two-dimensional case of subcritical
bifurcation, we find oscillating, stable, localized structures whose
time dependence is shown in Figure 2 in a series of snapshots.
This structure is robust and we have seen it with a large range of values
of the CGLE coefficients, emerging from a wide variety of
initial conditions.

Such oscillating pulses resemble the oscillons observed in recent
experiments of vertically-shaken layers of granular material$^{12,13,14}$,
and they also have emerged from other pattern equations$^{15,16,17,18}$.
We have here used the term oscillon to characterize the similar object
found with the subcritical CGLE.  The `-on' ending normally smacks of
integrability  and it may be that, in the astronomical context, the word
spicule would be more apt.  This is a matter for later discussion.

\section{Discussion}
Stratified layers with thermal dissipation frequently suffer acoustic
instabilities for various reasons that have not all been clarified as yet.
We have considered one of the simplest of these instabilities to show
what they may lead to.  In fact, it is well known that overstabilities in
thin layers where the spectrum of horizontal wavenumbers is so dense as
to be regarded as continuous have an amplitude function, or envelope, that
satisfies an equation of the CGLE form
when the amplitudes are not too large.  As we shall discuss elsewhere, the
inclusion of the effects of magnetic fields or rotation may not require
much qualification of these remarks.  So we are inclined to seek
applications of these considerations to thin layers such as stellar
chromospheres or slabs and disks.  Global modes such as are familiar in
helioseismology, being discrete, would require simpler treatments and
call for ODEs for their nonlinear description.  At any rate, we have been
able to extract from this simple theory solitary structures of a kind that
have been attributed to magnetic effects in discussions of solar
atmospheric dynamics.

It is not even necessary that the acoustic waves be unstable for a
description by the CGLE to be appropriate, as mechanical forcing could
also be included in the theory.  However, if the degree of instability or
the amplitude of the forcing becomes too large, one may question the use
of weakly nonlinear theory.  The recently observed excitation by a solar
flare of an expanding wave on the solar surface$^{19}$ is a case in 
point and it remains to be seen whether simple pattern equations can be
appropriate in such situations where the initial amplitudes are quite
large. Still, it is also true that the wave amplitude decays quickly
when the general background is stable and even then the weakly nonlinear
theory may be brought to bear.

L.T. thanks NSF for a Postdoctoral Fellowship 
and Pierre Coullet for a helpful conversation.


\newpage

\section{Figure Captions}

\noindent{Figure 1. Dispersion relation of the fundamental acoustic mode
($\gamma = 1.28$, $m = 1.5$, $z_0 = 0.1$ and $q = 3$).  The curves are
frequency and growth rate as a function of the horizontal wave number.}

\noindent{Figure 2. Evolution of an oscillon (CGLE parameters:
$\alpha = 1.0$, $\beta = 3.0 + i$ and $\eta = -2.75 + i$).
The domain is periodic in both spatial dimensions
with a period of $10\pi$.
The shaded surfaces are the real parts of the amplitude.  The
oscillation starts from the top left panel and continues to the right.}
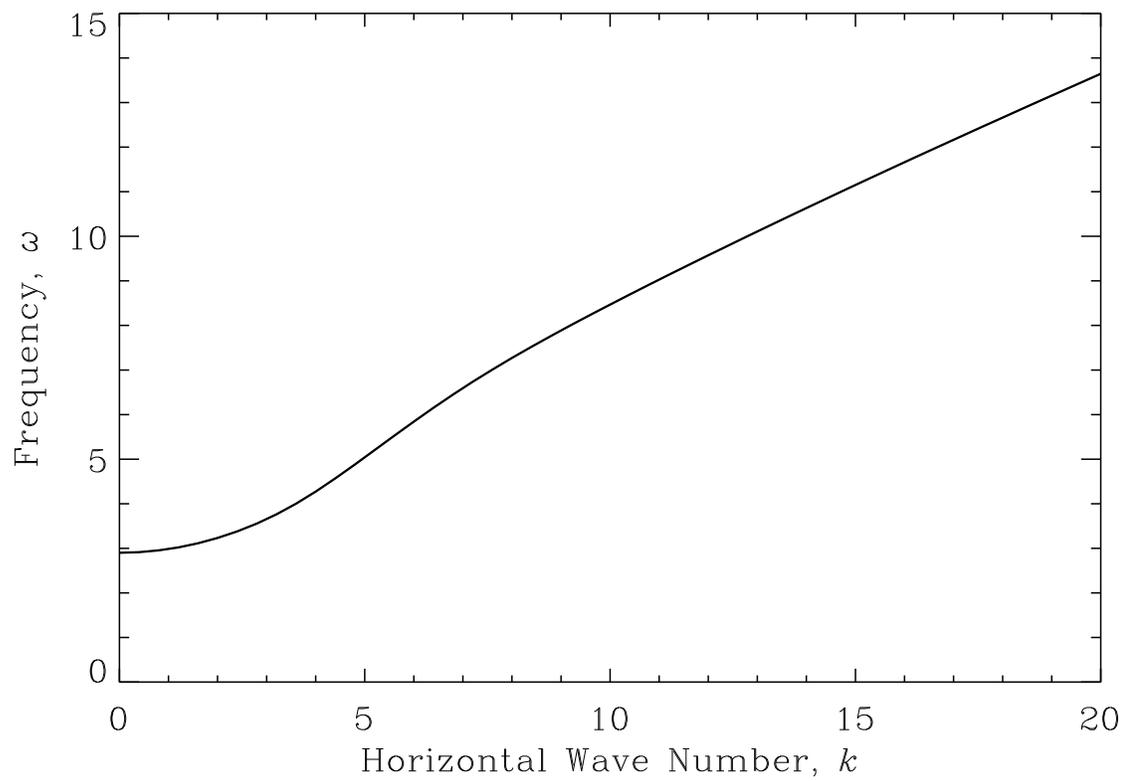
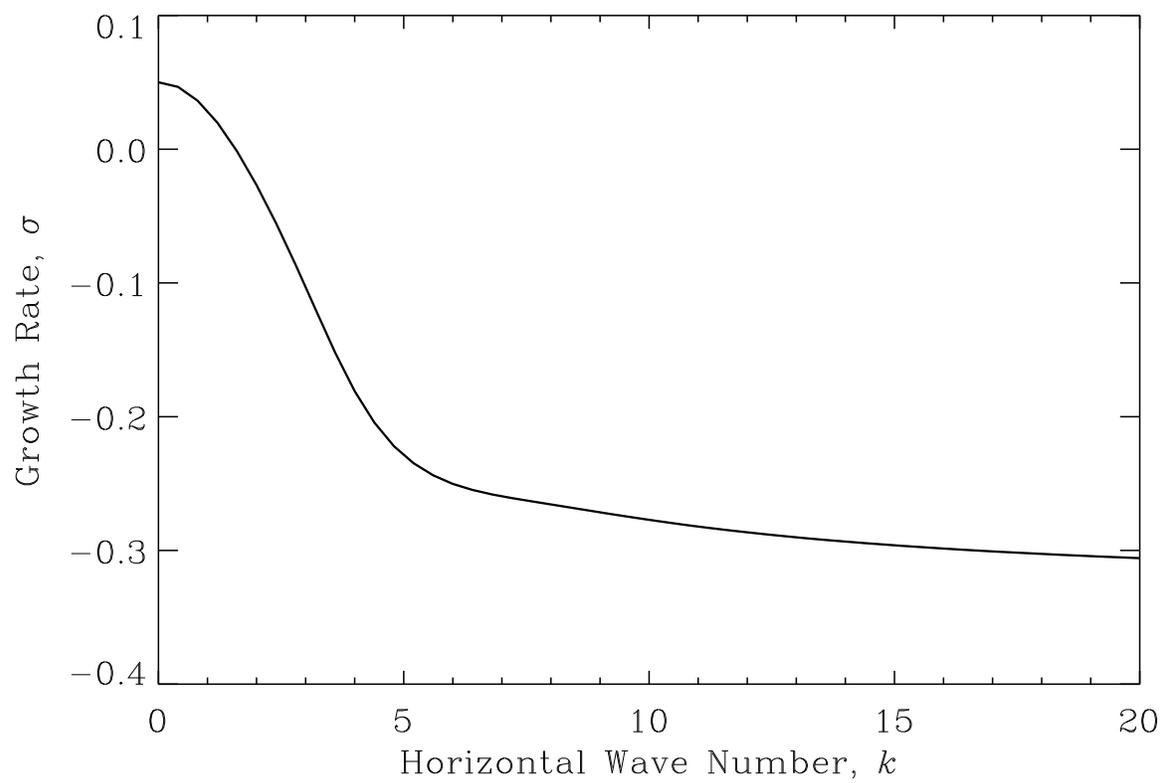
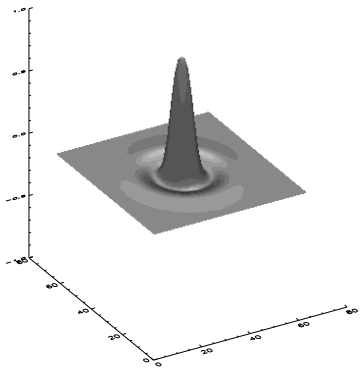
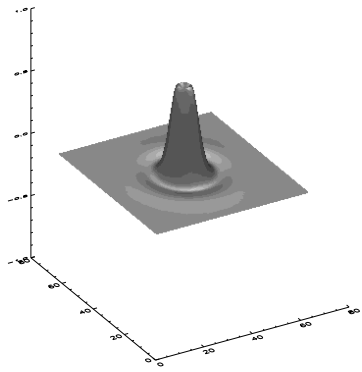
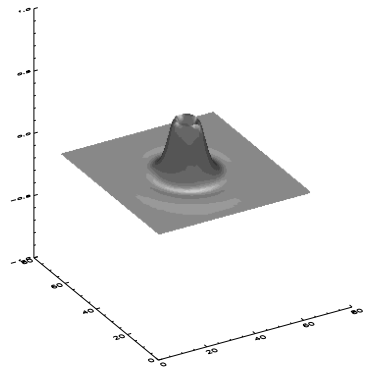
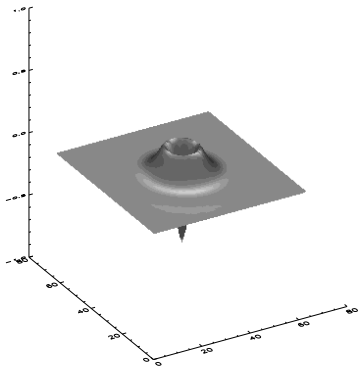
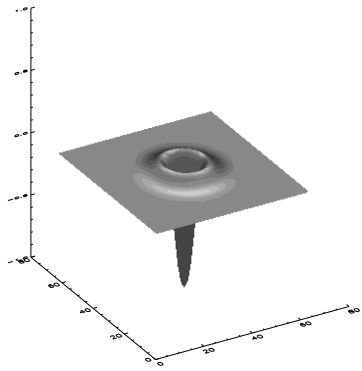
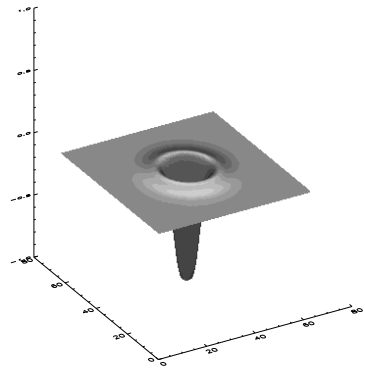
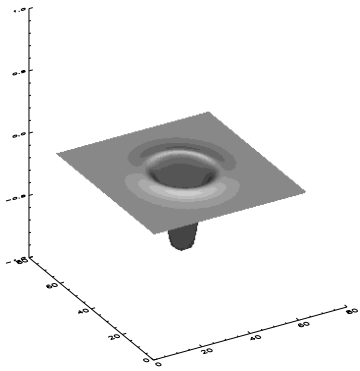
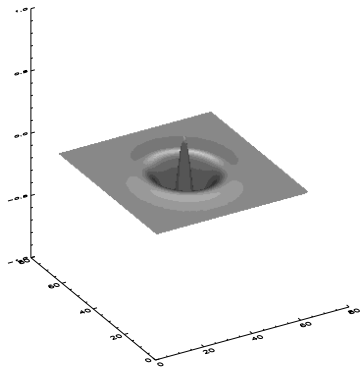
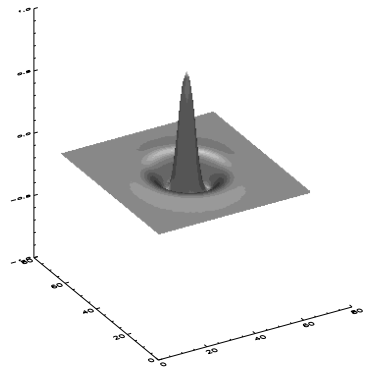
\end{document}